# Controllable anomalous Nernst effect in an antiperovskite antiferromagnet

Yunfeng You[1], Wenxuan Zhu[1], Hua Bai[1], Yongjian Zhou[1], Lei Han[1], Leilei Qiao[1], Tongjin Chen[1], Feng Pan[1] & Cheng Song[1]

[1]Key Laboratory of Advanced Materials, School of Materials Science and Engineering, Tsinghua University, Beijing 100084, China

Anomalous Nernst effect (ANE), the generation of a transverse electric voltage by a longitudinal temperature gradient, has attracted increasing interests of researchers recently, due to its potential in the thermoelectric power conversion and close relevance to the Berry curvature of the band structure. Avoiding the stray field of ferromagnets, ANE in antiferromagnets (AFM) has the advantage of realizing highly efficient and densely integrated thermopiles. Here, we report the observation of ANE in an antiperovskite noncollinear AFM $Mn_3SnN$ experimentally, which is triggered by the enhanced Berry curvature from Weyl points located close to the Fermi level. Considering that antiperovskite $Mn_3SnN$ has rich magnetic phase transition, we modulate the noncollinear AFM configurations by the biaxial strain, which enables us to control its ANE. Our findings provide a potential class of materials to explore the Weyl physics of noncollinear AFM as well as realizing antiferromagnetic spin caloritronics that exhibits promising prospects for energy conversion and information processing.

Introduction

Anomalous Nernst effect (ANE) refers to the phenomenon that a transverse electric voltage can be generated by a temperature gradient in a ferromagnetic conductor, which is perpendicular to both the magnetization and the temperature gradient. Conventionally, ANE has been considered to be proportional to the magnetization and thus observed only in ferromagnets [1–5]. Recent studies have shown that ANE can also be realized in chiral noncollinear antiferromagnet (AFM) $Mn_3Sn$ and $Mn_3Ge$ with zero or vanishingly small magnetization due to the topologically nontrivial Berry curvature at the Fermi energy $E_F$ [6–10]. From a fundamental perspective, ANE in AFM is intriguing since it can be exploited as a sensitive probe for determining key topological parameters of the Weyl physics [7, 8]. From a technological perspective, researching ANE of AFM is promising to design highly efficient and densely integrated thermopiles since the lateral configurations of thermoelectric modules enable us to enhance the coverage of heat source without suffering from the stray fields that can not be avoided in ferromagnets [7, 9].

With abundant magnetic and structural transitions, antiperovskite manganese nitrides $Mn_3AN$ (where A represents Sn, Ni, Ga, etc., owing a perovskite crystal structure but the locations of anion and cation are interchanged), have attracted a revival of attention due to various novel physical phenomena, such as negative thermal expansion [11, 12], baromagnetic [13], barocaloric [14], and piezomagnetic effects [15–17]. Lately, spin-orbit torque switching and unconventional spin polarizations have been investigated in these system [18–22], showing tremendous application potential in the field of spintronics. For magnetotransport behavior, the discovery and analysis of anomalous Hall effect (AHE) in AFM antiperovskites like $Mn_3SnN$ and $Mn_3NiN$ unravel the underlying distinctive physics of these materials [23–25]. Theoretically, AHE originates from the integration of the Berry curvature for all of the occupied bands, while ANE is decided by the Berry curvature at $E_F$. A large AHE thus can not guarantee a large ANE, and the ANE measurement should be a more sensitive probe for the Berry curvature near $E_F$ than AHE which probes the whole Fermi sea [7, 8, 26-28]. There has been theoretical research on the topology analysis of electronic structures of AFM antiperovskites [29], and large ANE has been predicted in the system [30]. However, the experimental observation of ANE in this class of materials has not been reported yet. Besides, compared with noncollinear AFM $Mn_3X$ systems (where X=Ge and Sn), the tunable noncollinear magnetic orderings of antiperovskites offer a convenient platform to modulate ANE by perturbations, such as temperature, doping and strain [13, 23, 25, 31, 32].

Here, we consider antiperovskite AFM $Mn_3SnN$ [the crystal structure is exhibited in Fig. 1(a)], where AHE has been both theoretically and experimentally proved to exist due to the Berry curvature from the noncollinear AFM structure [23, 29, 33]. In $Mn_3SnN$, there are two triangular AFM configurations, $\Gamma_{5g}$ (Fig. 1(b)) and $\Gamma_{4g}$ (Fig. 1(c)), where only the $\Gamma_{4g}$ structure can induce AHE [23, 29, 33]. Similarly, from the group theory analysis of how the spin orders impact on the Berry curvature, ANE disappears by symmetry in the $\Gamma_{5g}$ phase [30]. Moreover, since strain could induce the transition of magnetic phases in antiperovskites or influence the magnetic symmetry of the triangular configurations, the Berry curvature and ANE can be affected accordingly, like the case of AHE [23, 25, 30]. In this paper, we use $Mn_3SnN$ (001) films to research the ANE of AFM antiperovskites. The existence of Weyl points near $E_F$ of $Mn_3SnN$ is responsible for the observed ANE. We also find that ANE can be effectively manipulated by introducing different biaxial strain. Our research proposes a new class of materials to explore the Weyl physics of noncollinear

AFM. The ANE in Mn$_3$SnN is also promising to realize antiferromagnetic spin caloritronics [34, 35] that has bright prospects for energy conversion and information processing.

Methods

50 nm-thick (001)-oriented Mn$_3$SnN films were deposited by magnetron sputtering on MgO (001) substrates in an Ar+N$_2$ mixture gas atmosphere using a Mn$_{81}$Sn$_{19}$ target. Different biaxial strain was introduced by controlling the N$_2$ gas partial pressure during the growth of films. The growth of Mn$_3$SnN films on MgO (001) substrates with biaxial strain ε being 1.77%, 0.87%, and –0.26% has been investigated in our previous work [23], where the negative value denotes the compressive strain and the positive value denotes the tensile biaxial strain. Then, an amorphous 50 nm-thick Al$_2$O$_3$ insulating layer was deposited at 150 ℃ by atomic layer deposition (ALD). Finally, 10 nm-thick Pt layer was deposited on the Al$_2$O$_3$ insulating layer by magnetron sputtering at room temperature, providing a thermal gradient along the *z* direction of the device as the heater.

X-ray diffraction (XRD) and X-ray reflection (XRR) of the Mn$_3$SnN films were measured by Cu Ka1 radiation with *λ* being 1.5406 Å. The surface roughness was characterized by the atomic force microscope. Magnetic properties were measured using the superconducting quantum interference device (SQUID) magnetometry at different temperatures with the field up to 5 T. The magnetotransport measurements of ANE were conducted using a physical property measurement system (PPMS). The magnetic field is up to 6 T.

The energy band of Mn$_3$SnN was obtained by the first-principles calculations, using the Vienna *ab initio* simulation package (VASP) [36, 37] with projector augmented wave method [38, 39]. Perdew-Burke-Ernzerhof (PBE) functional [40] was used to treat the exchange correlation interaction and the plane-wave basis. The Gamma centered k-point mesh of 8 ×8 ×8 was used in all calculations.

Thermal simulations were carried out with finite element method (FEM) using COMSOL Multiphysics software kit. To estimate the temperature difference between the upper and lower surfaces of the $Mn_3SnN$ layer, two-dimensional simulations were carried out on a cross-section of the sample. To determine the steady state temperature distribution of the system with a non-zero current density in the Pt layer, the heat equation $\nabla \cdot (\kappa \nabla T) + Q = 0$, was deployed in the FEM simulations, where $T$ is the temperature field to be determined, and $Q$ is the Joule heat source term. In simulations, a Dirichlet boundary condition was given to the bottom surface of the substrate. The temperature of the bottom surface was set to the ambient temperature, $T_{amb} = 100$ K. On the left and right edges of the substrate, periodic boundary conditions were assigned. The rest edges which stand for surfaces directly exposing to the atmosphere had the thermally insulating condition, with the normal heat flux density $q_n = 0$. Since all layers are deposited with tight combination, interfacial thermal resistances were omitted for simplicity. The volume heat source in the Pt layer was determined through Joule's law, $Q_{Pt} = I^2 R/V_{Pt}$, where $V_{pt}$ stands for the volume of the Pt layer deposited in experiments [10, 41, 42].

Results and discussion

To study ANE of $Mn_3SnN$, we first check the energy band of $Mn_3SnN$ with the magnetic configuration of $\Gamma_{4g}$ from the first-principles calculations (see Fig. S1 within the Supplemental Material) [43]. To see the band structure near the Fermi energy clearly, Fig. 1(d) exhibits the energy band with ± 0.4 eV of the plane T including R.

We can observe that near the Fermi energy, there is a pair of Weyl points (marked by the red circles), which is also verified in the Ref. [29]. The existence of Weyl points around $E_F$ provides a precondition for the appearance of ANE in Mn$_3$SnN.

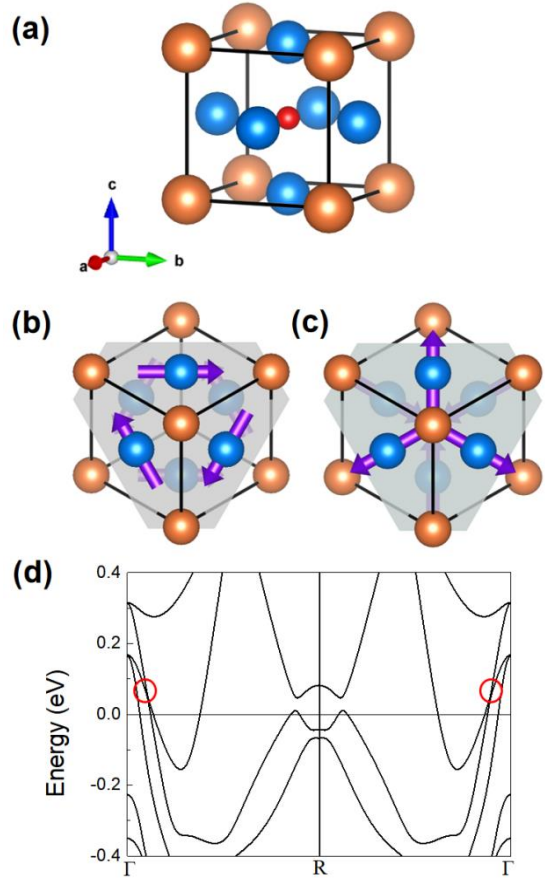

FIG. 1. (a) Crystal structure of Mn$_3$SnN, where the blue, orange and red spheres represent the Mn, Sn and N atoms, respectively. Schematics of the magnetic structure of Mn$_3$SnN (b) $\Gamma_{5g}$, and (c) $\Gamma_{4g}$, where purple arrows denote magnetic moments of Mn atoms and the gray plane donates the Kagome (111) plane. (d) The energy band of Mn$_3$SnN with the magnetic configuration of $\Gamma_{4g}$ obtained by the first-principles calculations. (e) The band structure around Fermi energy of Mn$_3$SnN. A pair of Weyl points are marked by the red circles.

Out-of-plane XRD patterns for 50 nm Mn$_3$SnN thin films grown on MgO

substrates with different biaxial strain $\varepsilon$ (1.77%, 0.87%, and −0.26%) are shown in Fig. 2(a). Apart from the diffraction peaks from the MgO substrate, we can see clear diffraction peaks of Mn$_3$SnN (00$l$) planes, indicating the quasi-epitaxial growth of the (001)-oriented Mn$_3$SnN film. Besides, within the sensitivity of XRD measurements, there is no secondary phase. To determine the in-plane orientation and the epitaxial relationship of the Mn$_3$SnN film, Φ scans are carried out for the Bragg peaks of the Mn$_3$SnN film and the MgO substrate. The inspection of Φ-scan (see Fig. S2 within the Supplemental Material) [43] shows that the peaks are separated by 90°. The fourfold symmetry for both the film and the substrate confirms the crystallographic orientation relationship being MgO(001)[100]//Mn$_3$SnN(001)[100]. The XRD patterns also exhibit Mn$_3$SnN (002) peaks with different positions, revealing different $c$ lattice constant of the films. Therefore, different biaxial strains are introduced into the film ascribed to the volume change of the lattice, and can be calculated using the formula $\frac{\Delta c/c_0}{\varepsilon} = -\frac{2\upsilon}{1-\upsilon}$, where $\Delta c/c_0$ represents the relative variation in $c$-axis length and $\upsilon$ is the Poisson ratio [15-17, 23]. The surface morphology of the film measured by atomic force microscope shows that the whole film is continuous and flat (see Fig. S3 within the Supplemental Material) [43]. The surface roughness $Ra$ is 0.278 nm for the film with biaxial strain being 0.87%.

The magnetic property of the film is characterized by SQUID, where we apply the in-plane magnetic field, consistent with the measurement configuration for the ANE curves. Fig. 2(b) presents three magnetization curves of the (001)-oriented Mn$_3$SnN films with different biaxial strain at 100 K. At a glance, all curves show a diamagnetic behavior due to the diamagnetic background of the MgO substrate, indicating the AFM characteristic of the Mn$_3$SnN (001) films. Corresponding magnetization curves after subtracting the diamagnetic background are presented in Fig. 2(c). Remarkably, the films with biaxial strain being 0.87% and −0.26% display a tiny in-plane uncompensated magnetization of ~1.81 and ~2.65 emu cm$^{-3}$, respectively, and the magnetization of the film with biaxial strain being 1.77% reaches ~17.3 emu cm$^{-3}$ (the larger magnetization is due to the piezomagnetic effect of Mn$_3$SnN [15–17, 23]). The appearance of uncompensated magnetization may be caused by the imperfect growth or the slight spin canting of the film, which is also verified in other noncollinear AFMs [23, 24, 44, 45].

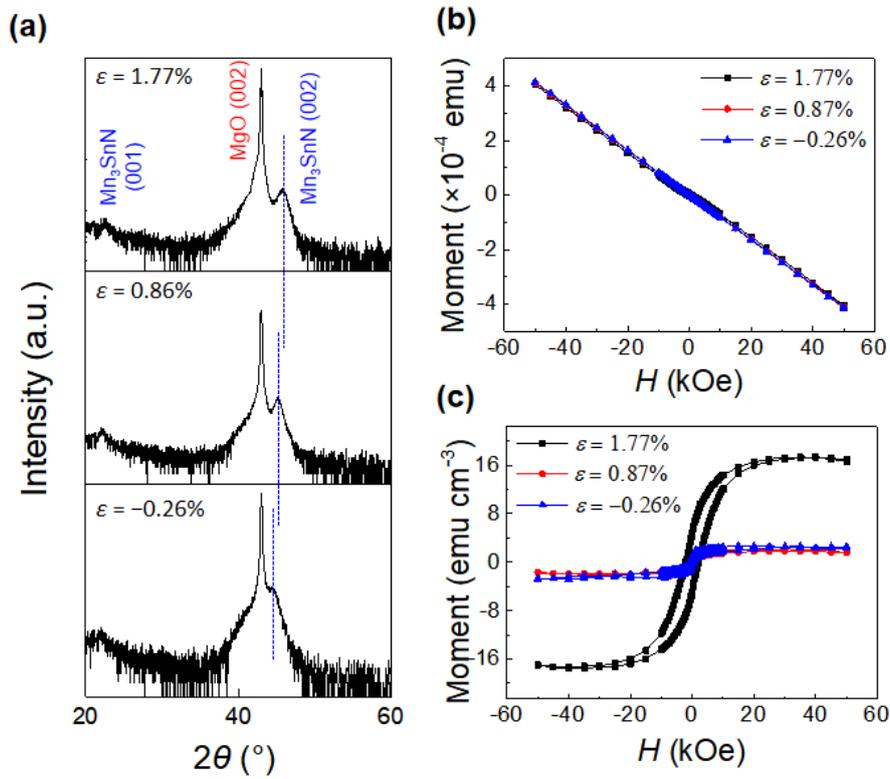

FIG. 2. (a) XRD patterns of the (001)-oriented Mn$_3$SnN films deposited on MgO (001) substrate with different biaxial strain $\varepsilon$. Magnetization curves of the Mn$_3$SnN (001) films with different biaxial strain $\varepsilon$ by applying an in-plane magnetic field (b) with and (c) without diamagnetic background from the substrate at 100 K.

Then, we measure the ANE of the (001)-oriented Mn$_3$SnN film with different biaxial strain, using the measurement geometry shown in Fig. 3(a) at 100 K. The temperature gradient $\nabla T$ is applied out-of-plane along the $z$ direction by the Pt heater, and the applied magnetic field is in-plane along the $y$ direction. The anomalous Nernst voltage $V_{ANE}$ is thus measured, perpendicular to both the temperature gradient and external magnetic field. Fig. 3(b) presents the magnetic field-dependent $V_{ANE}$ (the applied electrical current for heating is 15 mA) for (001)-oriented Mn$_3$SnN film with biaxial strain being −0.26%, 0.87%, and 1.77%, respectively. The film with $\varepsilon = -0.26\%$ merely shows a linear relationship of $V_{ANE}$ and the magnetic field, revealing that $\Gamma_{5g}$ dominates the magnetic configuration of Mn$_3$SnN in this case [30]. For Mn$_3$SnN film with $\varepsilon = 0.87\%$, $V_{ANE}$ reverses sign when the magnetic field is swept between ± 6 T and the reversal is hysteretic. The observation of spontaneous, zero field value reveals obvious ANE in Mn$_3$SnN, indicating that the magnetic configuration is mainly $\Gamma_{4g}$ [30]. Considering the tiny magnetization of the film, the obvious ANE signals observed in our experiments are mainly originated from the Berry curvature of the Weyl points around the Fermi energy as shown in Fig. 1(d). Besides, as exhibited in Fig. 2(c), the above two films have comparable magnetization, further confirming that the clear hysteresis loop of ANE signals is not caused by the uncompensated magnetization [7-9]. ANE is enhanced in the film with larger biaxial strain ($\varepsilon = 1.77\%$), since strain may reduce the magnetic space group of $\Gamma_{4g}$ magnetic configuration from $R\bar{3}m'$ to $C'2/m'$ [23, 24, 29, 33, 46] through the canting of the magnetic moments, further enhancing the Berry curvature of Mn$_3$SnN. The existence of $\Gamma_{4g}$ magnetic configuration here is verified by comparing the in-plane and out-of-plane magnetization of the film (see Fig. S4 within the Supplemental Material).

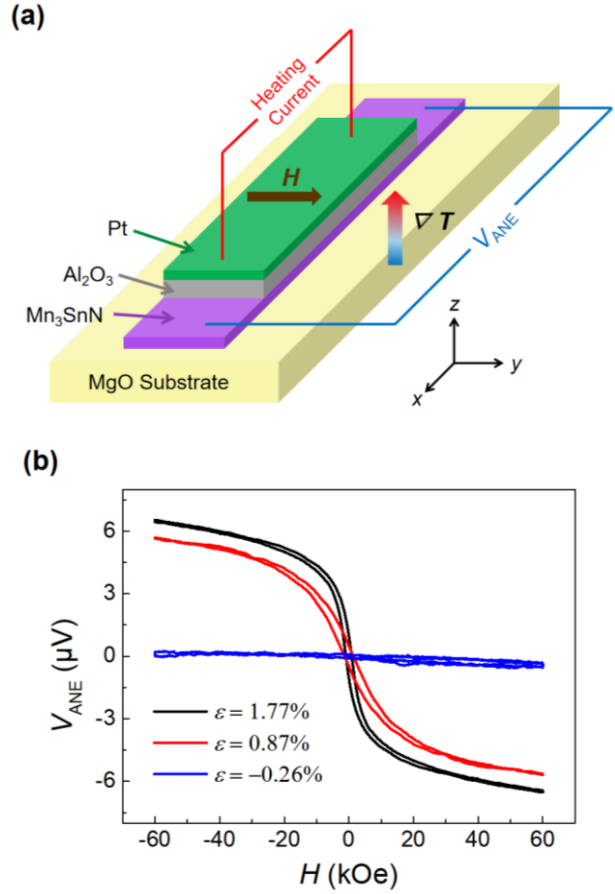

FIG. 3. (a) Schematic of the measurement of anomalous Nernst effect. The temperature gradient $\nabla T$ is applied out of plane along the $z$ axe by the Pt heater. The magnetic field is applied in plane along the $y$ axe. The anomalous Nernst voltage $V_{ANE}$ is measured along the $x$ axe. (b) Magnetic field dependence of $V_{ANE}$ for films with different biaxial strain $\varepsilon$ measured at 100 K. The applied electrical current for heating is 15 mA.

To further understand the observed ANE signals, we measure the field dependence of $V_{ANE}$ in $Mn_3SnN$ films with different electrical current $I$. The electrical current applied to the Pt heater is in the range from 1 mA to 15 mA. As shown in Fig. 4(a), for film with biaxial strain being 0.87%, obvious ANE signals can be observed with electrical current lager than 1 mA, and $V_{ANE}$ increases with the increasing electrical current. Likewise, the same phenomenon can be observed in film with biaxial strain being 1.77% as well (see Fig. S5 within the Supplemental Material) [43]. Accordingly, the dependence of $V_{\text{ANE}}$ derived from the ANE on $I^2$ of the above Mn$_3$SnN films with different biaxial strain is shown in Fig. 4(b). We can observe that $V_{\text{ANE}}$ is proportional to $I^2$, which satisfies the definition of ANE, since Joule heat thus $\nabla T$ produced by the Pt layer is proportional to $I^2$. From the slope of the lines, we can conclude that Mn$_3$SnN film with $\varepsilon = 1.77\%$ has the largest transverse anomalous Nernst coefficient $S_{\text{xz}}$, according to the formula

$$S_{\text{xz}} = \frac{V_{\text{ANE}}}{\Delta T} \times \frac{L_{\text{z}}}{L_{\text{x}}} \qquad (1)$$

where $L_{\text{x}}$ and $L_{\text{z}}$ represent the lengths of Mn$_3$SnN layer along the $x$ and $z$ (thickness) axes.

To confirm that the measured voltage signal is originated from ANE, we also use negative electrical current in the Pt layer, as shown in Fig. 4(c). Evidently, the ANE curves are almost coincident under both positive and negative current, independent of the current direction in the Pt heater. This phenomenon reveals the typical feature of ANE because whether positive or negative current in the Pt heater generates $\nabla T$ along the same direction. For comparison, we measure AHE of the film with both positive and negative electrical current flowing through Mn$_3$SnN at 100 K (the detection current is 1 mA). Differently, the sign of anomalous Hall Voltage $V_{\text{AHE}}$ changes with the polarity of the electrical current in the Mn$_3$SnN film as shown in Fig. 4(d). Therefore, the hysteresis loop in Fig. 4(a) arises from ANE rather than other effects.

To obtain the value of the anomalous Nernst coefficient of Mn$_3$SnN, we first need to acquire the temperature difference between the top and bottom of Mn$_3$SnN film. Since the Joule heat produced by the Pt heater defuses along the $z$ direction of the device, and the electrical contact of Mn$_3$SnN and Pt heater is separated by the Al$_2$O$_3$ insulator, the direct detection is difficult. Hence, the temperature difference $\Delta T$ here is evaluated by simulations. The thermal power density produced by Pt is obtained by measuring the actual resistance of the Pt layer. The geometrical size of the device is 1000 μm × 100 μm. The material properties that are needed for the simulation of the temperature distribution are the solid heat capacity at constant pressure $C_p$, the solid thermal conductivity $\kappa$ and the solid density. $C_p$ of Mn$_3$SnN is 75 J (mol K)$^{-1}$ [47] and the solid density is 8.47 g cm$^{-3}$. Since Mn$_{3.1}$Zn$_{0.5}$Sn$_{0.4}$N is also an antiperovskite manganese nitride with similar crystal structure to Mn$_3$SnN, the solid thermal conductivity of Mn$_3$SnN was assumed using $\kappa$ of Mn$_{3.1}$Zn$_{0.5}$Sn$_{0.4}$N, being 11 W (m K)$^{-1}$ [48]. Through the finite element modeling simulation, the temperature distribution of the schematic model along the $z$ axe (thickness direction) with $I$ being 15 mA in the Pt heater is shown in Fig. 4(e). Results of the simulations show that the most significant temperature drop takes place within the Al$_2$O$_3$ layer, and the temperature difference $\Delta T$ for the upper and lower surfaces of Mn$_3$SnN layer is calculated to be 6.9×10$^{-3}$ K. Therefore, according to the Eq. (1), the anomalous Nernst coefficient $S_{xz}$ of Mn$_3$SnN film with $\varepsilon$ of 0.87% and 1.77% is 0.08 and 0.09 μV/K, respectively. The value of $S_{xz}$ in our Mn$_3$SnN film is comparable to other noncollinear AFM film like Mn$_3$Ge thin film (0.10 μV/K) [9], and is smaller than 0.3 [7] or 0.27 [10] μV/K observed in Mn$_3$Sn single crystal. The relatively small value of our experiments is possibly due to the mix of $\Gamma_{4g}$ and $\Gamma_{5g}$ structures along with the imperfect growth of the thin films. Besides, like AHE [21, 22, 49], the maximum ANE can be obtained when the magnetic field is applied in the Kagome plane, yet we are unable to access this measurement geometry using our (001)-oriented thin films, thus the measured value of $S_{xz}$ is inevitably smaller.

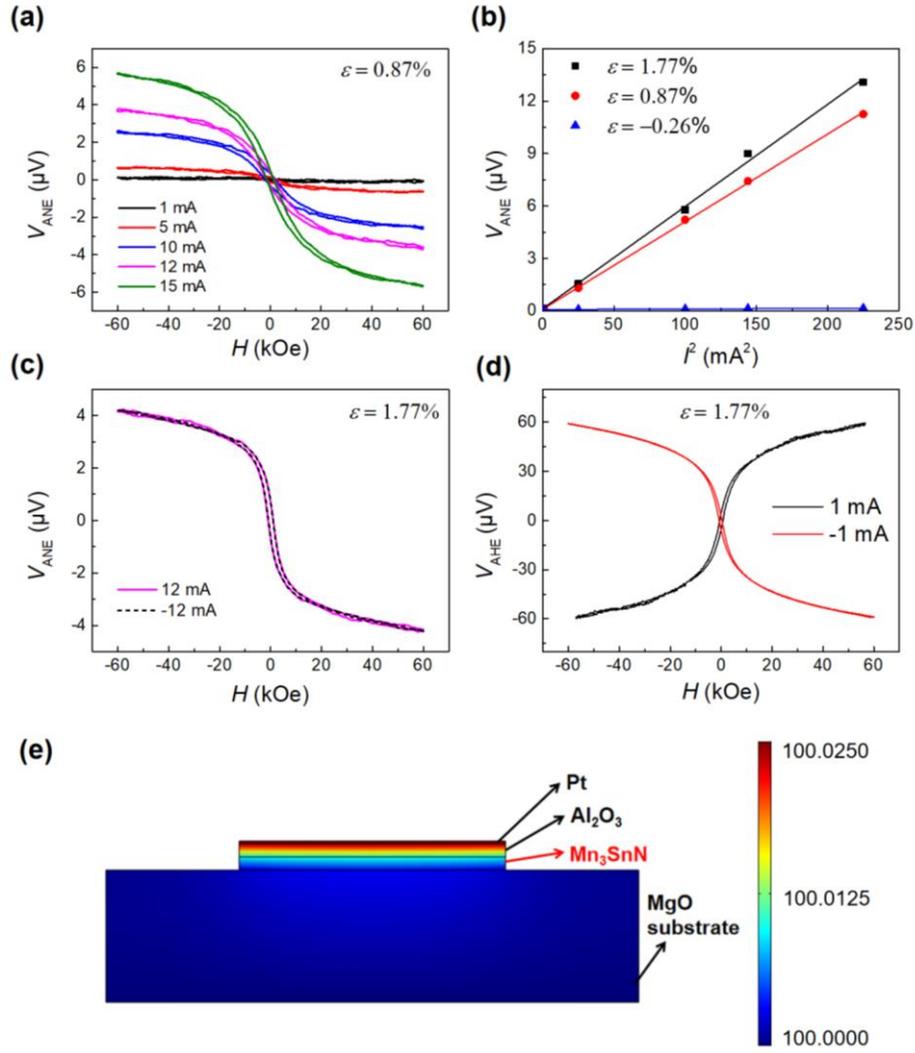

FIG. 4. (a) Field dependence of $V_{ANE}$ in Mn$_3$SnN film ($\varepsilon$ = 0.87%) with different electrical current $I$ at 100 K. (b) Dependence of $V_{ANE}$ on $I^2$ in the Pt heater in Mn$_3$SnN films with different biaxial strain. (c) ANE of Mn$_3$SnN film with biaxial strain being 1.77% measured under both positive and negative electrical current of 12 mA. (d) AHE of the same sample measured with the detection current of ± 1 mA. (e) Temperature distribution of the schematic model along the $z$ axe (thickness direction) using the finite element modeling simulation with $I$ = 15 mA in the Pt heater.

Conclusions

In conclusion, we have observed obvious ANE signals in the (001)-oriented

Mn$_3$SnN thin films at 100 K, which can be ascribed to the existence of Weyl points near $E_F$ of Mn$_3$SnN. ANE can be controlled by the biaxial strain in the film, which could induce the transformation of noncollinear AFM structures. Moreover, ANE can be further improved in films with larger biaxial strain, because strain may break the symmetry of the magnetic space group of $\Gamma_{4g}$ magnetic configuration through the canting of the magnetic moments due to the piezomagnetic effect of Mn$_3$SnN, thus, the Berry curvature of Mn$_3$SnN is influenced accordingly. On the one hand, our findings extend the material choice of ANE in AFM and provide a new class of materials to explore the Weyl physics of noncollinear AFM. On the other hand, controllable ANE in Mn$_3$SnN shows bright prospects for spin caloritronics that has extensive applications in energy conversion and information processing.